%% file: ms.tex
\documentclass{emulateapj}

\input{macros_vdbosch}

\begin{document}

\title{New Constraints on the Efficiencies of Ram-Pressure Stripping\\
and the Tidal Disruption of Satellite Galaxies}

\author{X. Kang\altaffilmark{1}, Frank C.  van den Bosch\altaffilmark{1}}

\altaffiltext{1}{Max-Planck-Institute for  Astronomy, K\"onigstuhl 17,
D-69117, Heidelberg, Germany}

\begin{abstract}
  Using data from the Sloan  Digital Sky Survey (SDSS) it has recently
  been  shown that the  red fraction  of satellite  galaxies increases
  with  stellar  mass. Semi-analytical  models,  however, predict  red
  satellite fractions  that are independent of stellar  mass, and much
  higher than observed. It has  been argued that this discrepancy owes
  to  the fact  that the  models  assume that  satellite galaxies  are
  instantaneously stripped  of their hot gas reservoirs  at the moment
  they are accreted  into a bigger halo.  In this  letter we show that
  the fraction  of red satellites  can be brought in  better agreement
  with  the  data by  simply  decreasing  this  stripping efficiency.  
  However, this  also results  in a red  fraction of  massive centrals
  that  is much  too low.   This  owes to  the fact  that the  massive
  centrals  now accrete  satellite galaxies  that are  bluer  and more
  gas-rich.  However, if a  significant fraction of low mass satellite
  galaxies is tidally disrupted before being accreted by their central
  host galaxy,  as suggested by  recent studies, the red  fractions of
  both centrals  and satellites can be reproduced  reasonably well.  A
  problem remains  with the red  fraction of centrals  of intermediate
  mass, which is likely to  reflect an oversimplified treatment of AGN
  feedback.
\end{abstract}

\section{Introduction}
\label{sec:intro}

Recently it has become   clear that semi-analytical models  for galaxy
formation   predict red satellite   fractions that are   much too high
(Weinmann \etal  2006; Baldry \etal  2006).  It  has been argued  that
this owes to  an oversimplified  treatment  of ram-pressure  stripping
(Gunn \&  Gott 1972) of  the hot gaseous  halos of satellite galaxies.
In all  semi-analytical models it is  assumed that the  entire hot gas
reservoir  of  a galaxy is  instantaneously stripped  the moment it is
accreted into a larger halo  (i.e., the moment  it becomes a satellite
galaxy).

Although  this was  originally motivated  by  the fact  that the  vast
majority  of  (faint)  satellite  galaxies  in clusters  are  red  and
deficient  in cold  gas  (e.g., Binggeli,  Tammann  \& Sandage  1987),
detailed analytical and hydrodynamical simulations have shown that the
typical  time  scale  for  this stripping  process  (sometimes  called
`strangulation') ranges  from $\sim 1$  Gyr to 10 Gyr  (e.g., Balsara,
Livio  \& O'Dea 1994;  Mori \&  Burkert 2000;  Bekki, Couch  \& Shioya
2002; Mayer \etal 2006; McCarthy \etal 2008).

In this letter we study the impact of a prolonged strangulation on the
colors of satellite and central  galaxies as function of stellar mass.
To that extent we implement a new recipe for ram pressure stripping of
hot gaseous halos of satellite galaxies in a semi-analytical model for
galaxy formation, and    compare the  red  fractions of    central and
satellite galaxies  to   data   obtained from  a  SDSS    galaxy group
catalogue.

\section{Semi-Analytical Model}
\label{sec:sam}

In order to model the colors of central  and satellite galaxies we use
the semi-analytical model of galaxy formation of Kang \etal (2005) and
Kang, Jing \&  Silk (2006; hereafter K06),  and we refer the reader to
these two papers for  details.  Briefly, dark  matter merger trees are
extracted  from a high  resolution $N$-body  simulation (with subhalos
resolved, Jing \&  Suto 2002) and populated  with galaxies following a
semi-analytical description.  Galaxies    are assumed to form  at  the
centers  of their dark matter  halos, using  recipes for cooling, star
formation, super-nova feedback, radio-mode  AGN feedback, and spheroid
formation that are  all very similar  to those of Croton \etal (2006).
When  a  dark matter halo   with its  associated `central'  galaxy  is
accreted  by  a larger halo, and  thus  becomes a subhalo, its central
galaxy becomes  a   `satellite'  galaxy.  The  orbital    evolution of
satellite galaxies  is tracked  using  the  orbital evolution of   its
corresponding subhalo, up to the point where  the subhalo is dissolved
in the simulation.  From  that  point on   the satellite galaxy   (now
called an `orphan  galaxy') is merged  with the central galaxy in  the
host halo after a dynamical friction  time computed using the standard
Chandrasekhar prescription.  Throughout this  paper, we adopt the same
model parameters as   in K06,  for  which  the  model reproduces   the
present-day   luminosity   function,  and   yields    a bimodal  color
distribution in  reasonable agreement with the data  (see below).  The
only changes we make here regard the treatment of the stripping of hot
gas around satellite galaxies.
\begin{figure}
\begin{center}
\includegraphics[width=1.0\columnwidth,height=0.75\columnwidth]{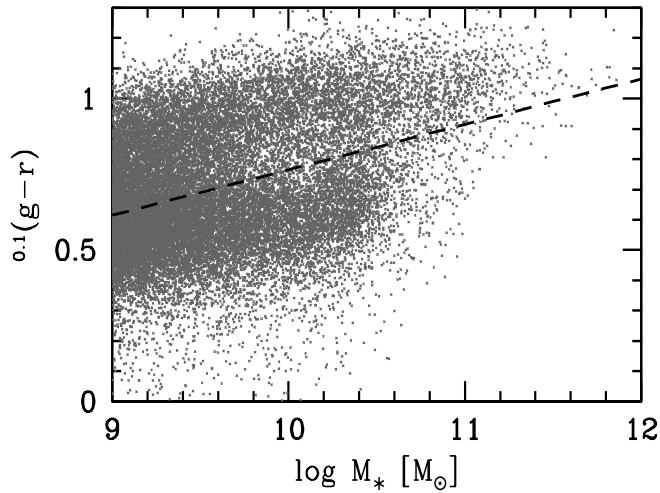}
\caption{The  distribution   of  model     galaxies as  function    of
  $^{0.1}(g-r)$ color and  the `observed' stellar  mass, obtained from
  the relation between stellar mass-to-light  ratio and color of  Bell
  \etal (2003).  The dashed line corresponds to  $^{0.1}(g-r) = 0.81 +
  0.15(\log[M_{\ast}/(h^{-2}M_{\odot})]-10)$ and  is used to split the
  model galaxies in red and blue populations (cf. Fig.7 in vdB07).}
\label{fig:cm}
\end{center}
\end{figure} 

In the standard model it is assumed that the hot gas associated with a
halo is instantaneously stripped as soon  as the halo is accreted by a
larger  halo  (i.e., as  soon  as  it  becomes a  subhalo).   However,
hydrodynamic simulations  indicate that the  ram-pressure stripping of
the hot gas  is not instantaneous, and may even  not be complete after
as much  as 10  Gyr.  Although the  characteristic time scale  of this
stripping process varies from one  simulation to the other, and ranges
from $\sim 1$  Gyr to 10 Gyr (see  \S\ref{sec:intro}), all simulations
suggest that the stripping  rate is approximately constant.  Motivated
by these findings  we model the rate at which  the subhalo is stripped
of its hot gas as
\begin{equation}\label{rate}
{{\rm d}M_{\rm hot} \over {\rm d}t} = -(1-f_{\rm hot})
{M_{{\rm hot},0} \over \tau}\,.
\end{equation}
Here $M_{{\rm hot},0}$ is the  hot halo gas mass  of the satellite  at
accretion, $\tau$ is the characteristic time  scale for stripping, and
$f_{\rm hot}$ is the fraction of hot gas left after one characteristic
time scale.  In addition to being stripped, the hot gas reservoir of a
satellite galaxy also decreases  due to cooling.   In order to compute
the cooling rate of the hot gas associated with a satellite galaxy, we
assume that the cooling radius and cooling  time of the hot gas remain
fixed at their values  at the time  of accretion.  We also assume that
the hot gas  remains associated with  the orphan galaxy even after its
subhalo is dissolved in the N-body simulation (we will discuss this in
\S\ref{sec:res}).  Finally, we   ignore any  potential  dependency  of
$f_{\rm hot}$ or $\tau$ on the satellite mass, the host halo mass, the
orbit  of  the subhalo,  or the  density  distribution of the  hot gas
associated    with the host  halo.  These    assumptions are certainly
oversimplified.  However,  the main aim of this  paper is  not to give
the  most accurate  description of the  rate at  which the hot  gas of
satellite galaxies is stripped, but merely  to investigate how changes
in $\tau$ and  $f_{\rm  hot}$ impact on the    colors of central   and
satellite galaxies.

\section{results}
\label{sec:res}

In a  recent paper, van den  Bosch \etal (2007;  hereafter vdB07) used
the SDSS galaxy group catalogue  of Yang \etal (2007) to determine the
red fractions of  central and satellite galaxies as  function of their
stellar masses.  Here  central galaxies have been defined  as the most
massive  group  members (in  terms  of  their  stellar masses),  while
satellite galaxies are all group members that are not centrals.  Using
the fact  that the  color-stellar mass relation  of SDSS  galaxies is
clearly bimodal, vdB07 split the  galaxies in blue and red populations
using the stellar-mass dependent cut,
\begin{equation}\label{colcut}
^{0.1}(g-r) = 0.76 + 0.15 \left( \log[M_{\ast}/(h^{-2}M_{\odot})]-10\right)\,.
\end{equation}
Here  $^{0.1}(g-r)$ is  the  $(g-r)$ color  $K$-corrected to  redshift
$z=0.1$,  and  $M_{\ast}$  is  the  stellar mass  computed  using  the
relation between  stellar mass-to-light ratio and color  of Bell \etal
(2003; eq.~[2] in Yang \etal 2007).

For our model galaxies, we compute the  $^{0.1}(g-r)$ colors using the
stellar  population synthesis code of Bruzual  \& Charlot (2003) for a
Salpeter IMF\footnote{We  have  verified   that our results   are  not
  sensitive   to  the choice  of  the  IMF.}.  In  order  to model the
photometric errors  in  the SDSS, we add  a  random, Gaussian error of
0.05 magn  to both the $^{0.1}g$  and $^{0.1}r$-band magnitudes of our
model  galaxies.  Finally,   for   each model galaxy   we  compute the
`observed'  stellar  mass  using  the  same  relation  between stellar
mass-to-light ratio and color as  used for the data. The color-stellar
mass relation thus obtained for  our fiducial model  (with $\tau = 0$,
corresponding to  instantaneous stripping of the hot  gas) is shown in
Fig.~\ref{fig:cm}. It clearly shows a bimodal color distribution, with
a red sequence  that extends to   more massive galaxies  than the blue
sequence. This owes mainly to our inclusion  of AGN feedback (see also
Croton  \etal 2006 and Bower \etal  2006), and is in good, qualitative
agreement with the data (cf. Fig.~7 in vdB07). Upon closer inspection,
though, we find that the colors of the red sequence and the bimodality
scale are $\sim 0.05$ magn redder than for the  SDSS. It is unclear at
present what the origin of this color offset is. In order to limit its
impact on our  analysis, we split our model  galaxies in red  and blue
populations using  eq.~(\ref{colcut}) but  with  a zero-point  that is
$0.05$ magn redder (indicated by the dashed line in Fig.~\ref{fig:cm})
\begin{figure*}
\begin{center}
  \includegraphics[scale=0.9]{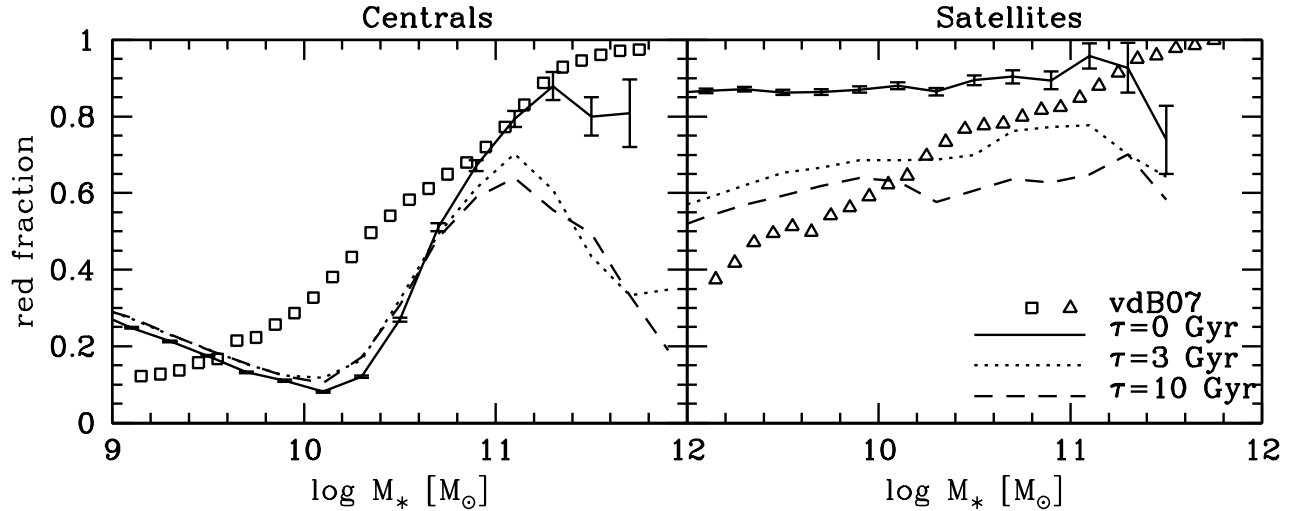}
  \caption{The fraction   of  red   centrals  (left-hand    panel) and
    satellites (right-hand panel) as  function of stellar mass.   Open
    symbols correspond to the results of vdB07, obtained from the SDSS
    group  catalogue   of  Yang \etal   (2007), while  different lines
    correspond to the results obtained  from our semi-analytical model
    for  different values   of the   characteristic   time  scale  for
    strangulation, as  indicated.  Errorbars   are obtained  using the
    jackknife  technique and,  for  clarity,  are  only shown  for the
    fiducial model with instantaneous stripping (i.e, $\tau=0$)}
\label{fig:fred}
\end{center}
\end{figure*}

Fig.\ref{fig:fred}  compares  the  red  fractions, $f_{\rm  red}$,  of
centrals  (left-hand  panel)  and  satellites  (right-hand  panel)  as
function  of  `observed'  stellar  mass obtained  from  our  fiducial,
instantaneous stripping  model (solid line) with the  results of vdB07
(open symbols). Note that this model predicts a red satellite fraction
that is $\sim 0.85$, almost  independent of stellar mass, and in clear
disagreement with the data.   As demonstrated in Weinmann \etal (2006)
and  Baldry \etal (2006)  the semi-analytical  models of  Croton \etal
(2006)  and  Bower  \etal  (2006),  which  also  assume  instantaneous
stripping of  the hot gas, suffer  from exactly the  same problem.  As
for the centrals, the model  roughly predicts the correct red fraction
at  the  massive  end,  which  is  mainly  due  to  the  inclusion  of
radio-model AGN  feedback.  However, for centrals  with $M_{\ast} \lta
10^{11}  \Msun$, the red  fraction predicted  by the  model is  a poor
match to the data. At the low mass end ($M_{\ast} < 10^{10} \Msun$),
it is upwards and inconsistent  with the data. These low-mass galaxies
are  too  red because  the  cold gas  are  consumed  rapidly in  their
earlier,  low-mass progenitors.   Observational work  (Kennicutt 1998)
shows that stars form efficiently only in galaxies with surface density
higher than a threshold, so  star formation should be more inefficient
in low-mass progenitors,  but this process  is ignored in  our model.
Another disagreement  is seen at $M_{\ast} \sim  10^{10} \Msun$, where
model galaxies are  too blue. We will discuss  this in \S\ref{sec:concl}
and in this  section we focus on the model  predictions at the massive
end ($M_{\ast} \sim 10^{11} \Msun$).

As  discussed   in  \S\ref{sec:intro},  the   over-prediction  of  red
satellites  could  be due  to  the  inaccurate  treatment of  hot  gas
stripping  from satellite  galaxies.  To  test this,  we  now consider
models with  less efficient stripping. Seen  from eq.(\ref{rate}) that
this can be  obtained by increasing $\tau$ or  $f_{\rm hot}$, and both
are found to  have similar trends on model  predictions.  Here we show
only  results by  changing $\tau$.   The  dotted and  dashed lines  in
Fig.~\ref{fig:fred} show  the results obtained  using eq.~(\ref{rate})
with $f_{\rm  hot}=0.3$ and $\tau =  3$ Gyr and  10 Gyr, respectively.
The effects  of a prolonged stripping  are twofold.  First  of all, as
expected,  it reduces the  red fraction  of satellite  galaxies.  Note
that  there is relatively  little difference  between the  models with
$\tau =  3$ Gyr and 10 Gyr,  and that neither model  can reproduce the
detailed stellar mass dependence of $f_{\rm red}$ observed.  Secondly,
both models predict  a much lower red fraction  among the most massive
central  galaxies,  ruining  the  good  agreement  achieved  with  the
fiducial model.  This  owes to the fact that  the central galaxies now
`accrete'  more blue  satellite galaxies  with significant  amounts of
fresh cold  gas which triggers a  starburst in the  central galaxy. In
principle,  we  could improve  the  fit of  the  red  fraction of  the
satellites at the  low mass end by making  $f_{\rm hot}$ and/or $\tau$
depend on  the stellar  mass of the  satellite galaxy, but  this would
only worsen the red  fraction of the (massive) centrals.  Furthermore,
since  the massive centrals  are now  too blue  (on average),  the red
fraction of  the massive satellites  (which were centrals  before they
were accreted)  is now also  too low.  Clearly, simply  prolonging the
stripping of the  hot gas of satellite galaxies  creates more problems
than it solves.

\begin{figure*}
\begin{center}
\includegraphics[width=1.8\columnwidth,height=1.15\columnwidth]{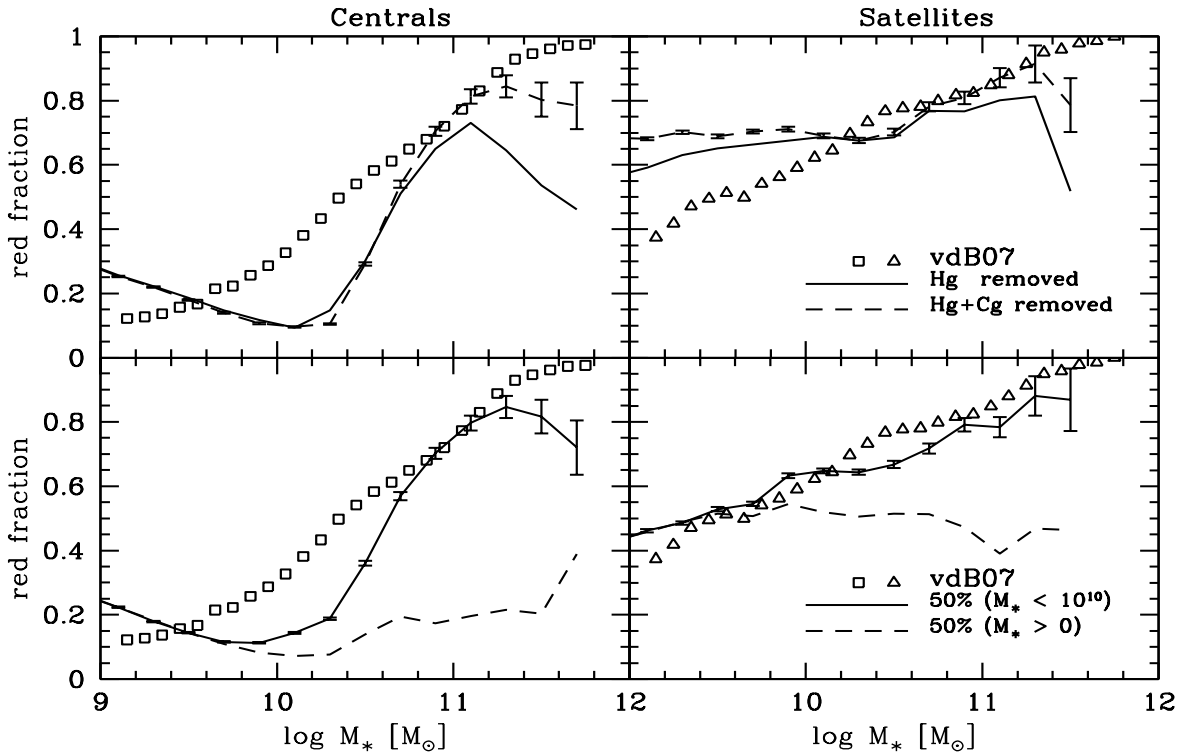}
\caption{Same as   Fig.~\ref{fig:fred}, but showing   the results  for
  models in  which we change certain properties  of satellite galaxies
  once they  become orphans (i.e., once their  associated subhalos are
  dissolved  in  the N-body  simulation).   In  the  upper panels  the
  reservoirs  of hot  gas  (solid lines)  and  both hot  and cold  gas
  (dashed  lines)  are  removed  once the  satellite  galaxies  become
  orphans.  In the  lower panels it is assumed that  50 percent of all
  satellite galaxies with $M_{\ast}  < 10^{10}\Msun$ (solid lines) and
  with $M_{\ast} >  0$ (dashed lines) are tidally  disrupted once they
  become orphans.  All  models shown here model the  hot gas stripping
  of  satellite galaxies using  eq.~(\ref{rate}) with  $\tau=3$Gyr and
  $f_{\rm hot}=0.3$.}
\label{fig:fred1}
\end{center}
\end{figure*}

How can  we  prevent the   massive galaxies  from   becoming too blue?
Clearly,  we need  to  prevent  the  (massive) central galaxies   from
accreting orphan galaxies that are too  blue and that have significant
amounts of cold  gas. As a first   attempt we assume that  a satellite
galaxy is completely stripped of its hot gas once it becomes an orphan
galaxy (i.e., when its subhalo in the numerical simulation dissolves).
The resulting   red fractions are shown   as solid lines  in the upper
panels of Fig.\ref{fig:fred1}. A comparison  with the dotted lines  in
Fig.~\ref{fig:fred}    shows that this  has  little  impact on the red
fractions.  If we assume, however, that the satellite is also stripped
of its cold gas once it becomes an orphan  we obtain the results shown
by  the dashed     lines in Fig.\ref{fig:fred1}.     Although in  good
agreement  with the data  at the  massive  end, we don't consider this
model very physical.

Recently, a number  of studies have argued that  the lack of evolution
in  the massive  end  of the  galaxy  stellar mass  function, and  the
presence of a significant  amount of intra-cluster light, suggest that
a significant  fraction of  satellite galaxies is  (tidally) disrupted
before it has the chance of  being accreted by its central host galaxy
(e.g.  Mihos et al.  2005; Zibetti  et al.  2005; Monaco et al.  2006;
Conroy,   Wechsler   \&  Kravtsov   2007).    The   lower  panels   of
Fig.\ref{fig:fred1} show the red  fractions obtained if we assume that
a  certain fraction of  satellite galaxies  is tidally  disrupted once
they become orphans. The solid lines correspond to a model in which 50
percent of all  orphans with a stellar mass  $M_{\ast} < 10^{10}\Msun$
are tidally disrupted.  This model yields red satellite fractions that
are  in  good agreement  with  the  data,  without causing  a  serious
under-prediction  of  the  red  fraction  of  massive  centrals.   For
comparison, the  dashed lines correspond to  a model in  which 50\% of
{\it all} orphan galaxies  are tidally disrupted, independent of their
stellar  mass.  This  causes  a drastic  under-prediction  of the  red
fractions of massive centrals and  satellites, which owes to fact that
the  number  of major  mergers  experienced  by  a central  galaxy  is
effectively reduced  by a factor  two.  Consequently, the  majority of
massive  centrals  are  now  disk  dominated and  the  associated  AGN
feedback (assumed to be proportional  to the mass of the spheroid, see
K06) is not sufficiently efficient to suppress the cooling of hot gas.

\section{Conclusions}
\label{sec:concl}

We have used a semi-analytical  model of galaxy formation to study the
fractions of red centrals and satellites as functions of their stellar
mass.  In  agreement with Weinmann  et al.  (2006)  and Baldry et  al. 
(2007) we find that the  models severely over-predict the red fraction
of satellite  galaxies.  This owes  to an oversimplified  treatment of
ram-pressure  stripping  of  the  hot gas  associated  with  satellite
galaxies.  In  particular, all semi-analytical models  assume that the
hot gas reservoir  of a galaxy is instantaneously  stripped the moment
it  is accreted  into a  bigger halo  (i.e., the  moment it  becomes a
satellite  galaxy).  We have  demonstrated that  the fractions  of red
satellites can  be brought in better  agreement with the  data, if the
efficiency  of  this  stripping  process  is  significantly  reduced.  
However, the implication is that massive centrals also become bluer to
the extent that  their red fraction is much too low.  This owes to the
fact that the satellite galaxies that are accreted by the centrals are
now bluer and more gas rich.  This problem can be prevented if roughly
half  of  all satellite  galaxies  with  a  stellar mass  $M_{\ast}  <
10^{10}\Msun$  are tidally disrupted  prior to  being accreted  by the
central host galaxy. This is in qualitative agreement with the lack of
evolution in the massive end  of the galaxy stellar mass function, and
with the presence of a significant amount of intra-cluster light.

All models presented  here predict a red fraction  of central galaxies
with $M_{\ast}  \sim 10^{10} \Msun$ that is  significantly too low.
As shown by Baldry et al.  (2006), the semi-analytical model of Croton
et al.  (2006) suffers from  exactly the same problem, while the model
of Bower et al. (2006) seems  to be in better agreement with the data.
The  main difference  between the  model of  Bower et  al.   and those
presented here and in Croton et al. is the way in which the efficiency
of AGN feedback  scales with galaxy properties. In  fact, for galaxies
with $M_{\ast}  \sim 10^{10} \Msun$, QSO activities  are commonly seen
across cosmic  time, and it  has been shown that QSO wind can  also quench
their star  formation (e.g., Di  Matteo,  Springel  \&
Hernquist 2005).  As already eluded  to in Weinmann et al.  (2006), it
thus seems that  the red fractions of central  galaxies as function of
their stellar  mass yield important  constraints on the  efficiency of
AGN  feedback  (or  any  other  mechanism that  can  quench  the  star
formation  in  central galaxies,  e.g.,  Birnboim,  Dekel \&  Neistein
2007). Although  a study of  radio AGN or  QSO feedback is  beyond the
scope  of this  paper, we  emphasize that  a modification  of  the red
fraction of central galaxies will  also modify that of the satellites.
After all,  the former are the  direct progenitors of  the latter.  In
particular,  a model  that  would  predict a  higher  red fraction  of
centrals with $M_{\ast} \sim  10^{10}M_{\odot}$ requires an even lower
efficiency of ram pressure stripping of the hot gas of satellites than
obtained here.

\end{document}

%% file: macros_vdbosch.tex
\newcommand{\etal}{{et al.~}}

\newcommand{\Msun}{\>{\rm M_{\odot}}}

\newcommand{\beq}{\begin{equation}}
\newcommand{\eeq}{\end{equation}}

\def\gtsima{$\; \buildrel > \over \sim \;$}
\def\ltsima{$\; \buildrel < \over \sim \;$}
\def\prosima{$\; \buildrel \propto \over \sim \;$}
\def\gsim{\lower.7ex\hbox{\gtsima}}
\def\lsim{\lower.7ex\hbox{\ltsima}}
\def\simgt{\lower.7ex\hbox{\gtsima}}
\def\simlt{\lower.7ex\hbox{\ltsima}}
\def\simpr{\lower.7ex\hbox{\prosima}}
\def\la{\lsim}

\def\lta{\la}

\newdimen\hssize
\newdimen\hdsize
